\documentclass{PoS}

\title{Cosmic Polarization Rotation in view of the recent CMB experiments}

\ShortTitle{Cosmic Polarization Rotation}

\author{\speaker{Sperello di Serego Alighieri}\\
        INAF - Osservatorio Astrofisico di Arcetri, Firenze, Italy\\
        E-mail: \email{sperello@arcetri.astro.it}}

\abstract{The possibility that there is some Cosmic Polarization Rotation (CPR), i.e. that the polarization angle rotates while a photon travels in vacuum over large distances, is important for at least two reasons: first, the polarization angle seems to be the most steady characteristic of photons and, second, CPR would be associated with violations of fundamental physical principles, like the Einstein Equivalence Principle on which all metric theories of gravity are based, including General Relativity, for which we celebrate the Centennial this year 2015.
We review here the astrophysical tests which have been carried out to check if CPR exists.
These are using the radio and ultraviolet polarization of radio galaxies and the polarization of
the cosmic microwave background (both E-mode and B-mode). These tests so far have been negative, leading to upper limits of 
a couple of degrees on any CPR angle, thereby increasing our confidence in those physical principles and in the resulting theories,
including General Relativity.
We discuss future prospects in detecting CPR or improving the constraints on it.
}

\FullConference{XI Multifrequency Behaviour of High Energy Cosmic Sources Workshop -MULTIF15-,\\
		25-30 May 2015\\
		Palermo, Italy}

\begin{document}

\section{The context}

In this year 2015 we celebrate two important events, relevant for this paper: the International Year of Light, promoted by UNESCO together with many scientific and engineering societies \footnote{see http://www.light2015.org/Home.html}, and the Centennial of the theory of General Relativity (GR), developed by Einstein in 1915 \cite{Ein15}. The photons, i.e. the \textit{quanta} of light, are very important for us astronomers, since they are the carrier of almost all the information that we have about the Universe, outside the Solar System (the only exceptions are a few cosmic rays and several elusive neutrinos). The information carried by photons is parametrized by 1) their direction, as coded by celestial coordinates, 2) their energy, equivalent to the wavelength or frequency of the associated electromagnetic radiation, and 3) the position angle ($PA$) of their polarization. Is this information changed while photons travel in vacuum? The answer is definitely "yes" for the first two parameters, since the direction is changed by a strong gravitational field and the energy is changed by the expansion of the Universe. The Cosmic Polarization Rotation (CPR) deals with eventual changes of the polarization $PA$.

CPR is interesting also because, if it exists, it should be either positive, for a counter clockwise rotation (we follow the IAU convention for the polarization angle \cite{iau74}), or negative for a clockwise rotation. This immediately suggests that a non-zero CPR would be connected with the violation of fundamental physical principles, having to do with symmetry.
Indeed CPR is connected with Lorentz invariance violation, CPT violation, neutrino number asymmetry, and violation of the Einstein Equivalence Principle (EEP), on which GR is based \cite{Nio10}. In fact a null CPR greatly increases our confidence in GR. In this GR centennial year let us examine in some detail why CPR is connected with a test of the EEP. 

Since the Weak Equivalence Principle (WEP) is tested to a much higher accuracy than the EEP,\  Schiff \cite{Sch60} conjectured that any consistent Lorentz-invariant theory of gravity which obeys the WEP would necessarily also obey the EEP. If this were true, the EEP would tested to the same accuracy as the WEP, increasing our experimental confidence in GR. However Ni \cite{Nio73, Nio77} found a unique counter example to Schiff's conjecture: a pseudoscalar field which would lead to a violation of the EEP, while obeying the WEP.\  Such field would produce a CPR. Therefore testing for the CPR is important for our confidence in GR.

In the following I shall review the astrophysical tests of CPR, from the early ones involving radio galaxies (RG) to the more recent ones using the cosmic microwave background (CMB), show how they are all consistent with a null CPR, and discuss the prospects for improving the current results. A similar but more extensive review is available elsewhere \cite{diS15}.

\section{Early CPR tests using radio galaxies}

Testing for CPR is simple in principle: it requires a distant source of linearly polarized radiation, for which the orientation $PA_{em}$ of the polarization at the emission can be established. Then a CPR angle can be obtained by comparing the observed orientation $PA_{obs}$ with $PA_{em}$:
$$ \alpha = PA_{obs} - PA_{em}.$$
In practice it is not easy to know \textit{a priori} the orientation of the polarization for a distant source: in this respect the fact that scattered radiation is polarized perpendicularly to the plane containing the incident and scattered rays has been of great help, applied both to RG and to the CMB (see the next section). For those cases in which CPR depends on wavelength, one can also test CPR by simply searching for a variation of $PA_{obs}$ with the wavelength of the radiation, even without knowing $PA_{em}$.

The first test for CPR has been performed by looking at the difference between the $PA$ of the radio polarization, corrected for Faraday rotation, and the $PA$ of the radio axis of RG with $0.4< z < 2$ and linear polarization $P > 5\%$ \cite{Car90}. A peak in the distribution of the difference was found at $90 ^{\circ}$ and a smaller one at $0 ^{\circ}$. From the width of the distributions it was concluded that any rotation of the polarization should be smaller than $6.0$ $^{\circ}$ at the 95\% confidence level.

Re-examining the same data on the radio polarization of RG \cite{Car90}, a systematic rotation of the plane of polarization as large as 3 rad., independent of the Faraday one, and correlated with the angular position and with the distance of the RG, was claimed \cite{Nod97}.
However several authors \cite{War97, Eis97, Car97, Lor97} have independently and convincingly argued against this claim. In particular it was  shown that the misalignment for a sample of 29 quasars is consistent with zero and incompatible with the claimed rotation \cite{War97}.

However the test performed using the radio polarization of RG has some disadvantages: it requires a correction for the Faraday rotation, it is not based on a strict physical prediction of the polarization orientation at the emission, and holds only statistically on a large sample of sources. Therefore the test performed using the ultraviolet (UV) polarization of radio galaxies was a big improvement \cite{Cim94}. In fact the perpendicularity between the UV axis of every distant RG with $z > 0.5$ and with a polarization measurement, up to $z = 2.63$, and the direction of their plane of UV polarization showed that this plane is not rotated by more than $10$ $^{\circ}$. This perpendicularity is strictly expected for simple physical reasons, since the elongation and the polarization are due to scattering of anisotropic nuclear radiation: therefore it holds for every single RG, not just statistically on a large sample. Furthermore, Faraday rotation is completely negligible in the UV, so no correction is necessary. The method using the UV polarization of RG can be applied also to the polarization which is measured locally at any position in the elongated structures around the galaxy, and which has to be perpendicular to the vector joining the observed position with the nucleus. From the polarization map in the V-band ($\sim 3000$\AA \ rest-frame) of 3C 265, a radio galaxy at $z=0.811$ \cite{Tra98}, the mean deviation of the 53 independent polarization vectors from the perpendicular to a line joining each to the nucleus is $-1.4 ^{\circ} \pm 1.1 ^{\circ}$ \cite{War97}.

Recently we have perfomed an update of the CPR test using the UV polarization of RG \cite{diS10}. Using the polarization data for all 8 RG with $z > 2$, $P > 5\%$ in the UV ($\lambda \sim 1300 \AA$), and elongated UV morphology, we could show that, assuming uniform CPR, 
the average CPR angle at the mean redshift $\left\langle z \right\rangle =2.80$ must be: $$\alpha = -0.8 ^{\circ} \pm 2.2 ^{\circ}$$ which is well consistent with a null CPR.
Our data compilation \cite{diS10} has been used to set limits on a rotation of the plane of polarization, which depends on the direction in the sky  with a spherical-harmonic variation and a stochastic variation \cite{Kam10}. In the latter case the constraint is  $\left\langle \alpha ^2 \right\rangle \leq (3.7 ^{\circ})^2$. The CPR test using the UV polarization has advantages over the other tests at radio or CMB wavelengths, if CPR effects grow with photon energy (the contrary of Faraday rotation), as in a formalism where Lorentz invariance is violated but CPT is conserved \cite{Kos01, Kos02}.

In the mean time an important improvement has been suggested also for the CPR test using the radio polarization: since the radio emission is due to synchrotron, its Faraday-corrected polarization should be perpendicular to the projected magnetic field, which in turn is perpendicular to strong gradients in the radio intensity direction, as can be checked on high angular resolution radio data. For example, an average constraint on any CPR angle of $\alpha = -0.6 ^{\circ} \pm 1.5 ^{\circ}$ at the mean redshift $\left\langle z\right\rangle =0.78$ has been obtained \cite{Car98}, using the data on the radio polarization of 10 RG \cite{Lea97}.

\section{CPR tests using the Cosmic Microwave Background}

A more recent method to test for the existence of CPR is the one that uses the CMB polarization, which was induced by the last Thomson scattering of decoupling photons at $z\sim 1100$, resulting in a correlation between temperature gradients and polarization \cite{Lep98}. CMB photons are strongly linearly polarized, since they result from scattering. However the high uniformity of CMB produces a very effective averaging of the polarization in any direction. It is only at the CMB temperature disuniformities that the polarization does not average out completely and residual polarization perpendicular to the temperature gradients is observed, as expected. Therefore also for the CMB polarization it is possible to precisely predict the polarization direction at the emission and to test for CPR. After the first detection of CMB polarization anisotropies by DASI \cite{Kov02}, there have been several CPR tests using the CMB E-mode polarization pattern. 

Table 1 summarizes the most recent and accurate CPR measurements obtained using the CMB polarization. A more detailed version of this table is available elsewhere \cite{diS15}. The table also shows the systematic error deriving from a poor calibration of the polarization $PA$, which affects the CMB measurements because of the lack of suitable calibration sources at CMB frequencies and of the difficulties connected with an \textit{a priori} knowledge of the detector's orientation and with using ground-based calibration sources. The current $PA$ calibration accuracy is of the order of one degree, producing a non-negligible systematic error $\beta$ on the measured $PA$. In order to alleviate the $PA$ calibration problem, a self-calibration technique has been suggested \cite{Kea13}, consisting in minimizing EB and TB power spectra with respect to the $PA$ offset. Unfortunately such a calibration technique would eliminate not just the $PA$ calibration offset $\beta$, but $\alpha - \beta$, where $\alpha$ is the uniform CPR angle, if it exists. Therefore no independent information on the uniform CPR angle can be obtained, if this calibration technique is adopted, like with the BICEP2 \cite{Ade14b} and POLARBEAR \cite{Ade14c} experiments.

Another problem of CPR searches using the CMB is that unfortunately the scientists working on the CMB polarization have adopted for the polarization angle a convention which is opposite to the one used for decades by all other astronomers and enforced by the IAU \cite{iau74}: for the CMB polarimetrists, following a software for the data pixelization on a sphere \cite{Gor05}, the polarization angle increases clockwise, instead of counterclockwise, facing the source. This produces an inversion of the U Stokes parameter, corresponding to a change of $PA$ sign. Obviously, these different conventions have to be taken into account, when comparing data obtained with the different methods used for CPR searches. As an example of the problems raised, a review of CPR measurements was first published in a version which mixed results in the two opposite conventions, without taking them into account \cite{Kau15a}. Their figure 1 wrongly showed that most CPR measurements are on the negative side. The review was later corrected and the correct version of figure 1 actually shows a more even distribution of CPR measurements. Although the conclusion of the paper are not affected, since all CPR measurements are in any case consistent with zero, this mishap shows that the problem of using different conventions can be a serious one. As mentioned in the first section, all $PA$ in this work are given in the IAU convention: looking at the source, $PA$ increases counterclockwise. 

In summary, although some have claimed to have detected a rotation \cite{Xia10, Kau14}, the CMB polarization data appear well consistent with a null CPR. In principle the CMB polarization pattern can be used to test CPR in specific directions. However, because of the extremely small anisotropies in the CMB temperature and polarization, CMB tests so far have used averages over large regions of sky, assuming uniformity.

Recently constraints on the CPR have also been set using measurements of the B-mode polarization of the CMB, because of the coupling from E-mode to B-mode polarization that any such rotation would produce \cite{diS14}. This possibility is presently limited by the relatively large systematic errors on the polarization angle still affecting current data. The result is that from the South Pole Telescope polarimeter (SPTpol), POLARBEAR and BICEP2 B-mode polarization data it is only possible to set constraints on the fluctuations $\left\langle \delta\alpha ^2 \right\rangle \leq (1.56 ^{\circ})^2$ of the CPR, not on its mean value. Similarly an upper limit on the CPR fluctuations $\left\langle \delta\alpha ^2 \right\rangle \leq (1.68 ^{\circ})^2$ has been obtained \cite{Mei15} from the ACTPol B-mode data \cite{Nae14}. The last raw of Table 1 reports the combined constraint on the CPR fluctuations obtained from all the B-mode data mentioned above.

\begin{table} 

\begin{tabular}{@{}lccc@{}} 
Method & CPR angle $\pm$ stat. ($\pm$ syst.)& Frequency or $\lambda$ & Distance \\
\hline\noalign{\smallskip}
RG radio pol. & $\vert \alpha \vert < 6 ^{\circ}$ & 5 GHz & $0.4<z<1.5$ \\
RG UV pol. & $\vert \alpha \vert < 10 ^{\circ}$ & $\sim3000$ \AA\ rest-frame & $0.5<z<2.63$ \\
RG UV pol. & $\alpha = -1.4 ^{\circ} \pm 1.1 ^{\circ}$ & $\sim3000$ \AA\ rest-frame & z = 0.811 \\
RG radio pol. & $\alpha = -0.6 ^{\circ} \pm 1.5 ^{\circ}$ & 3.6 cm & $\left\langle z \right\rangle = 0.78$ \\
CMB pol. BOOMERanG & $\alpha = 4.3 ^{\circ} \pm 4.1 ^{\circ}$ & 145 GHz & $z \sim 1100$ \\
CMB pol. QUAD & $\alpha = -0.64 ^{\circ} \pm 0.50 ^{\circ} \pm 0.50 ^{\circ} $& 100-150 GHz & $z \sim 1100$ \\
RG UV pol. & $\alpha = -0.8 ^{\circ} \pm 2.2 ^{\circ}$ & $\sim1300$ \AA\ rest-frame & $\left\langle z \right\rangle = 2.80$ \\
RG UV pol. & $\left\langle \delta\alpha ^2 \right\rangle \leq (3.7 ^{\circ})^2$ & $\sim1300$ \AA\ rest-frame & $\left\langle z \right\rangle = 2.80$ \\
CMB pol. WMAP9 & $\alpha = 0.36 ^{\circ} \pm 1.24 ^{\circ} \pm 1.5 ^{\circ}$ & 23-94 GHz & $z \sim 1100$ \\
CMB pol. BICEP1 & $\alpha = 2.77 ^{\circ} \pm 0.86 ^{\circ} \pm 1.3 ^{\circ}$ & 100-150 GHz & $z \sim 1100$ \\
CMB pol. ACTPol & $\alpha = 1.0 ^{\circ} \pm 0.63 ^{\circ}$ $^{*}$ & 146 GHz & $z \sim 1100$ \\
CMB pol. B-mode & $\left\langle \delta\alpha ^2 \right\rangle \leq (1.36 ^{\circ})^2$ & 95-150 GHz & $z \sim 1100$ \\
\hline\noalign{\smallskip}

\end{tabular} 

$^{*}$A systematic error should be added, equal to the unknown difference of the Crab Nebula polarization $PA$ between 146 and 89.2 GHz.\\

\caption{Measurements of CPR with different methods (in chronological order).} 
\label{tab1} 

\end{table}

\section{Summary and outlook}

In the past 25 years CPR has been looked for with different methods, but none was found. Therefore the polarization PA appears to be the most steady property of photons, that are indeed able to trasmit this important geometrical information across the Universe. In practice all CPR test methods have reached so far an accuracy of the order of $1 ^{\circ}$ and $3\sigma$ upper limits to any rotation of a few degrees. It has been useful to use different methods, since they are complementary in many ways. They cover different wavelength ranges, and, although most CPR effects are wavelength independent, the methods at shorter wavelength have an advantage, if CPR effects grow with photon energy. They also reach different distances, and the CMB method clearly uses the longest traveling photons. However, the relative difference in light travel time between $z=3$ and $z=1100$ is only 16\%.
All methods can potentially test for a rotation which is not uniform in all directions; nevertheless this possibility has not yet been exploited by the CMB method, which also is not able to see how an eventual rotation would depend on the distance.
The dependence of CPR on the wavelength and on the distance of the source has been recently examined \cite{Gal14}, and none was found, which is not surprising for a null CPR, at least so far.

In the future improvements can be expected for all methods, e.g. by better targeted high resolution radio polarization measurements of radio galaxies and quasars, by more accurate UV polarization measurements of radio galaxies with the coming generation of giant optical telescopes \cite{deZ14, San13, Ber14}, and by future CMB polarimeters such as PLANCK \cite{Ade14a} and BICEP3 \cite{Ahm14}. Indeed the Planck satellite is expected to have a very low statistical error ($\sim 0.06 ^{\circ}$) for CPR measurements, about a factor of 10 better than previous experiments \cite{Xia08}. Unfortunately, although Planck has completed its observations about two years ago, its results on CPR have not yet been released. In any case, in order to exploit the much improved measurement accuracy of Planck, it will be necessary to reduce accordingly also the systematic error in the calibration of the polarization angle, which at the moment is of the order of $1 ^{\circ}$ at CMB frequencies. The best prospects to achieve this improvement are likely to be more precise measurements of the polarization angle of celestial sources at CMB frequencies, e.g. with the Australia Telescope Compact Array \cite{Mas13} and with ALMA \cite{Tes13}, and a calibration source on a satellite \cite{Kau15}.

\bigskip
\bigskip

\noindent{\bf Discussion}

\bigskip

\noindent{\bf Jim Beall:} There is a fair amount of evidences for large scale B-field structures in radio jets. Would these have any effect on your measurements? I wonder about possible optical emission from synchrotron processes.

\medskip

\noindent{\bf Sperello di Serego Alighieri:} Indeed there are a few cases of optical synchrotron emission from radio jets (e.g. M87 and 3C66B). In these cases the polarization direction is very similar in the optical and in the radio. However, the UV radiation of the powerful radio galaxies, which I am refering to, is completely independent of the radio emission and is thought to be scattered radiation from a quasar which is at the centre of the radio galaxy and emits anisotropically in two opposite cones, not in our direction.

\bigskip

\noindent{\bf Fr\'ed\'eric Marin:} In the UV polarization measurements of distant quasars, why don't you find both parallel and perpendicular polarization angle, if you have both face-on and edge-on objects? Is it due to a contribution from the jets?

\medskip

\noindent{\bf Sperello di Serego Alighieri:} The UV polarization measurements, which I have discussed, refer to the powerful radio galaxies, not to quasars. In such case the linear polarization $PA$ is expected to be perpendicular to the UV axis.

\end{document}